\documentclass[a4paper,11pt]{article}

\usepackage{float}
\usepackage[margin=2.5cm]{geometry}
\usepackage{graphicx}
\usepackage{numbertabbing}
\usepackage{times}
\usepackage{url}
\usepackage{xspace}

\usepackage[nodayofweek]{datetime}
\newdateformat{simple}{\THEDAY\ \monthname[\THEMONTH]\ \THEYEAR}
\simple

\usepackage{amsmath}
\allowdisplaybreaks[2]

\usepackage{amssymb}

\usepackage{amsthm}
\newtheoremstyle{plain-boldhead}
  {\topsep}
  {\topsep}
  {\itshape}
  {}
  {\bfseries}
  {.}
  { }
  {\thmname{#1}\thmnumber{ #2}\thmnote{ (\bfseries #3)}}
\newtheoremstyle{definition-boldhead}
  {\topsep}
  {\topsep}
  {\normalfont}
  {}
  {\bfseries}
  {.}
  { }
  {\thmname{#1}\thmnumber{ #2}\thmnote{ (\bfseries #3)}}
\theoremstyle{plain-boldhead}
\newtheorem{theorem}{Theorem}

\newtheorem{lemma}[theorem]{Lemma}
\newtheorem{corollary}[theorem]{Corollary}

\theoremstyle{definition-boldhead}
\newtheorem{definition}{Definition}

\newcommand\para[1]{\paragraph{#1.}}

\floatstyle{ruled}
\newfloat{algo}{htbp}{algo}
\floatname{algo}{Algorithm}

\newcommand{\N}{\mathbb{N}}

\newcommand{\false}{\str{false}\xspace}
\newcommand{\true}{\str{true}\xspace}

\newcommand{\str}[1]{\textsc{#1}}
\newcommand{\var}[1]{\textit{#1}}
\newcommand{\op}[1]{\textsl{#1}}
\newcommand{\msg}[2]{\ensuremath{[\str{#1}, {#2}]}}
\newcommand{\becomes}{\ensuremath{\leftarrow}}
\newcommand{\concat}{\ensuremath{\circ}}

\newcommand{\CO}{\ensuremath{\mathcal{O}}\xspace}
\newcommand{\CS}{\ensuremath{\mathcal{S}}\xspace}
\newcommand{\CR}{\ensuremath{\mathcal{R}}\xspace}
\newcommand{\CZ}{\ensuremath{\mathcal{Z}}\xspace}

\newcommand{\NAME}{COP\xspace}
\newcommand{\nil}{\str{null}}
\newcommand{\forklin}{fork-lin\-e\-ar\-iza\-bi\-lity\xspace}

\newcommand{\shortlist}{\parskip0pt\topsep0pt\partopsep0pt\parsep1pt\itemsep2pt}

\newcommand{\comment}[1]{}

\begin{document}

\newcommand{\thetitle}{Verifying the Consistency of Remote Untrusted
  Services with Conflict-Free Operations%
\footnote{A predecessor of this paper with a slightly different title
was presented at OPODIS 2014 and appears in the proceedings~\cite{cacohr14}.}
} 

\title{\bf \thetitle}
\date{8 February 2018}

\author{Christian Cachin$^1$\\
  IBM Research - Zurich\\
  \url{cca@zurich.ibm.com}
  \and Olga Ohrimenko$^2$\\
  Microsoft Research, Cambridge (UK)\\
  \url{oohrim@microsoft.com}
}

\maketitle

\footnotetext[1]{Corresponding author. IBM Research - Zurich,
  S\"{a}umerstrasse 4, CH-8803 R\"{u}schlikon, Switzerland.}

\footnotetext[2]{Work done at IBM Research~- Zurich and at Brown
  University.}

\begin{abstract}\noindent
  A group of mutually trusting clients outsources a computation
  service to a remote server, which they do not fully trust and that
  may be subject to attacks.  The clients do not communicate with each
  other and would like to verify the correctness of the remote
  computation and the consistency of the server's responses.
  This paper presents the \emph{Conflict-free Operation verification
    Protocol (\NAME)} that ensures linearizability when the server is
  correct and preserves \forklin in any other case.  All clients that
  observe each other's operations are consistent, in the sense that their
  own operations and those operations of other clients that they see are
  linearizable.  If the server forks two clients by hiding an operation,
  these clients never again see operations of each other.  \NAME supports
  \emph{wait-free} client operations in the sense that when executed with a
  correct server, \emph{non-conflicting} operations can run without waiting
  for other clients, allowing more parallelism than earlier protocols.  A
  conflict arises when an operation causes a subsequent operation to
  produce a different output value for the client who runs it.
  The paper gives a precise model for the guarantees of \NAME and includes
  a formal analysis that these are achieved.

\end{abstract}

\section{Introduction}

With the advent of \emph{cloud computing}, most computations run in
remote data centers and no longer on local devices.  As a result,
users are bound to trust the service provider for the confidentiality
and the correctness of their computations.
This work addresses the \emph{integrity} of outsourced data and
computations and the \emph{consistency} of the provider's respon\-ses.  
Consider a group of mutually trusting clients who want to collaborate on a
resource that is provided by a remote, partially trusted server.  This
could be a wiki containing data of a common project, an archival document
repository, or a groupware tool running in the cloud.  A subtle change in
the remote computation, whether caused inadvertently by a bug or
deliberately by a malicious adversary, may result in wrong responses to the
clients.  The clients do not trust the provider to always respond
correctly, hence, they would like to assess the integrity of the
computation, to verify that responses are correct, and to check that they
all get consistent responses.

In an asynchronous network model without communication among clients such
as considered here, a faulty or \emph{Byzantine} server may perform a
\emph{forking attack} and omit the effects of operations by some clients in
her responses to other clients.  Not knowing which operations other clients
execute, the forked clients cannot detect such violations.  The best
achievable consistency guarantee in this setting is captured by
\emph{\forklin}, introduced by Mazi{\`e}res and Shasha~\cite{mazsha02} for
storage systems.  It ensures that whenever the server in her responses to a
client~$C_1$ has ignored an operation executed by a client~$C_2$, then
$C_1$ can never again observe an operation by $C_2$ afterwards and vice
versa.  In other words, the views of the two clients remain forked after
the first inconsistency.  This property ensures clearly defined service
semantics in the face of an attack and allows clients to detect server
misbehavior easily.
For instance, the clients may periodically exchange a message
outside the model over a low-bandwidth channel and thereby verify the
correctness of a service in an end-to-end way.

Several conceptual~\cite{cashsh07,mdss09,cakesh09b,cakesh11} and practical
ad\-van\-ces \cite{wisish09,fzff10,mslcad11,scckms10,fbfw12} have already
improved consistency checking and verification with \forklin and related
notions.  These protocols ensure that when the server is correct, the
service is linearizable and ideally also \emph{wait-free}, that is, every
client's operations complete independently of other clients.  It has been
recognized, however, that \emph{conflicts} between operations can cause a
clients to block; this applies to fork-linearizable
semantics~\cite{mazsha02,cashsh07} and to other forking consistency
notions~\cite{cakesh09b,cakesh11}.  By weakening the notion of \forklin to
permit some inconsistent operations in the views of the forked clients, one
can circumvent blocking, as illustrated by FAUST~\cite{cakesh11} and
Ve\-nus~\cite{scckms10}.

In this paper, we go beyond storage services and address the consistency of
\emph{computation} coordinated by a Byzantine server.  The
\emph{Conflict-free Operation verification Protocol} or \emph{\NAME}
imposes fork-linearizable semantics for arbitrary functionalities and
allows clients to operate concurrently without blocking unless their
operations conflict.
\NAME extends earlier protocols aiming at the same goal, in particular, the
Blind Stone Tablet (BST) protocol~\cite{wisish09}; unlike previous works,
\NAME comes with a detailed formal analysis of its properties.

Supporting wait-free operations and avoiding server-side locks are key
features for efficient collaboration with remote coordination, as
geographically separated clients may operate at different speed.
Consequently, previous work has devoted a lot of attention to identifying
and avoiding blocking~\cite{mazsha02,cashsh07,limaz07}.  For example, two
read operations in a storage service never conflict.  On the other hand,
when a client writes a data item concurrently with another client who reads
it, the reader has to wait until the write operation completes; otherwise,
\forklin is not guaranteed~\cite{cashsh07}.  If \emph{all} operations are
to proceed without blocking, though, it is necessary to relax the
consistency guarantees to notions such as
weak \forklin~\cite{cakesh11}, for instance.  \NAME maintains the stronger
property of \forklin and always lets clients proceed at their own speed;
conflicting operations that would block are aborted, as considered by
Majuntke et al.~\cite{mdss09}.  The definition of conflicting operations in
\NAME is generic and corresponds to a ``write-read conflict'' between two
concurrent database transactions.

In \NAME, the server merely coordinates client-side operations but does not
compute the responses to operations nor maintain the service state.  This
conceptually simple approach can be found in many related
protocols~\cite{wisish09,hsgr10,fzff10} and practical collaboration systems
(such as \emph{git} or \emph{Mercurial} for source-code versioning).

\subsection{Contributions}

This paper considers a generic service executed by an untrusted server and
introduces the Conflict-free Operation verification Protocol (\NAME) with
the following properties:
\begin{itemize}\shortlist
\item \NAME provides wait-free, abortable consistency verification and
  ensures \forklin to a group of clients executing an arbitrary joint
  functionality and using a remote Byzantine server for coordination; it
  exploits sequences of non-conflicting operations.  The notion of conflict
  considered by \NAME corresponds to write-read conflicts in databases and
  generalizes commutative operations used in previous work.
\item \NAME comes with a detailed formal analysis and proof of correctness,
  showing that it achieves fork-linearizable semantics for generic service
  emulation; previous work did not establish this notion.
\end{itemize}
\NAME follows the general pattern of most previous
fork-lin\-e\-ar\-iz\-able emulation protocols, in particular the Blind
Stone Tablet (BST) protocol~\cite{wisish09}.  For determining when to
proceed with concurrent operations, we consider whether \emph{sequences} of
operations conflict and respect the state of the service, in contrast to
earlier protocols, which considered only isolated operations.

\NAME adopts the notion of conflict-freedom from VICOS~\cite{brcakn15}, which
appeared after the preliminary publication of \NAME.  VICOS also
illustrates one way to extend \NAME through authenticated data structures
such that the service state is held by the remote server instead of the
clients.  Combining \NAME with these appears possible but is left to future
work.

\subsection{Related work}
\label{ssec:related}

This section reviews related protocols according to their features and not
always in chronological order.  A summary of these systems in comparison
with \NAME is given in Table~\ref{tab:comparison}.

\para{Storage protocols}

Fork-linearizability has been introduced (under the name of \emph{fork
  consistency}) together with the SUNDR storage
system~\cite{mazsha02,lkms04}.  Conceptually SUNDR operates on storage
objects with simple read/write semantics.  Subsequent work of Cachin et
al.~\cite{cashsh07} improves the communication cost of untrusted storage
protocols to linear in the number of clients, compared to the quadratic
overhead of SUNDR.  A lock-free storage protocol (called \textsc{Concur})
was proposed by Majuntke et al.~\cite{mdss09} and introduced the idea of
aborting operations that might block.  This means that all operations
complete in the absence of contention, i.e., when they arrive one after the
other at the server; but with contention, all concurrent operations except
for one are aborted.  \textsc{Concur} does not distinguish between types
operations that conflict or not.

FAUST~\cite{cakesh11} and Venus~\cite{scckms10} offer only weak \forklin,
which excludes the last operation response from the consistency guarantee
(i.e., the last operation of one client may differ from the one seen by
another client and \forklin holds only after the client completes another
operation).  However, these two protocols also extend the model by
introducing occasional message exchanges among the clients.  This allows
FAUST and Venus to obtain stronger semantics, in the sense that they
eventually reach consistency (i.e., linearizability) or detect server
misbehavior.  In the more restrictive model considered here, \forklin is
the best possible guarantee~\cite{mazsha02}.

In Depot~\cite{mslcad11}, an even weaker notion called fork-causal
consistency is ensured for the stored data by the core algorithm.  Through
client-to-client communication, Depot also supports ``join'' operations
after forks and achieves a condition called fork-join-causal consistency.
This resembles the ``eventual consistency'' of geo-replicated cloud storage
systems, where views may fork temporarily and be reconciled again later.
In contrast, \NAME aims at the stronger notion of \forklin and does not
consider communication among clients.

All protocols mentioned so far except Depot use vector clocks (or even
vectors of vector clocks) for keeping track of the ordering relation among
operations.  Depot instead constructs a hash chain over all operations.

\para{Generic services}

\NAME builds on the Blind Stone Tablet (BST) protocol~\cite{wisish09},
which first extended \forklin consistency verification from storage to
generic services.  It considers a database hosted by a remote, untrusted
server and propagates state updates to all clients after they have been
ordered.  Every client builds a hash chain over the operation log for
keeping track of the consistency with other clients.

BST allows \emph{some} client operations that commute to proceed
concurrently and aborts others that do not commute.  However, it achieves
only limited wait-freedom, even for commuting operations.  Furthermore,
there is no formal analysis of the consistency achieved by the BST
protocol.  The discussion of fork-linearizability by Williams et
al.~\cite{wisish09} only addresses data\-base state updates, but not the
responses output by clients.  We elaborate on these shortcomings in
Section~\ref{ssec:bst} and provide in Section~\ref{sec:analysis} a detailed
analysis of \NAME as a key contribution of this work.

\newcommand{\Y}{\ensuremath{\checkmark}}

\begin{table*}
  \centering
  \begin{tabular}{|l||c|c|c|c|c|}
    \hline
    {\bf Protocol} & {\bf Wait-free} & {\bf Function} & {\bf Consistency} &
    {\bf Execution} & {\bf Proof} \\ \hline
    SUNDR~\cite{mazsha02,lkms04} & --- & storage & fork-lin. & server & --- \\ \hline
    FAUST~\cite{cakesh11}, Venus~\cite{scckms10} & \Y & storage & weak fork-lin. & server & \Y \\ \hline
    \textsc{Concur}~\cite{mdss09} & --- & storage & fork-lin. & clients & \Y \\ \hline
    BST~\cite{wisish09} & (\Y) & single commuting op.  & (fork-lin.) & clients & --- \\ \hline
    SPORC~\cite{fzff10} & \Y & generic op.~transform & (weak fork-lin.) & clients & --- \\ \hline
    Depot~\cite{mslcad11} & --- & storage & fork-join-causal & clients & \Y \\ \hline
    VICOS~\cite{brcakn15} & \Y & storage, conflict-free & fork-lin. & server & --- \\ \hline
    \NAME (this work) & \Y & generic, conflict-free & fork-lin. & clients & \Y \\ \hline
  \end{tabular}
  \caption{Summary of related protocols.
    In this table under \emph{function}, the BST protocol supports only
    a \emph{single} commuting operation and does not achieve wait-freedom
    (as indicated by the parentheses in the first column);
    SPORC is wait-free for generic functions that have \emph{operational
      transforms}; \NAME is wait-free for generic 
    \emph{non-conflicting} operation sequences.
    For the \emph{consistency} property, 
    \emph{weak~\forklin} (and \emph{\mbox{fork-*} linearizability}) allows 
    the last operation of a client to be inconsistent compared to 
    \emph{\forklin}; however, BST and SPORC do not guarantee their 
    consistency notion for client responses, only for state changes that may 
    occur much later (as indicated by the parentheses).
    The \emph{execution} column indicates whether the \emph{clients} compute
    operations and maintain state or if this is done by the \emph{server}.
    Finally, \emph{proof} indicates whether an algorithm has been formally 
    proven correct. Note that VICOS appeared after the initial publication 
    of \NAME.
  }
  \label{tab:comparison}
\end{table*}

\para{Non-blocking protocols}

In the context of cloud services replicated over wide-area networks, the
tradeoff between fault-tolerance, availability, and consistency has
received a lot of attention through the ``CAP Theorem''~\cite{brewer00}.
Services that explicitly allow operations to proceed in parallel whenever
possible provide an attractive way to circumvent the impossibility of being
``available'' and ``consistent'' simultaneously in the presence of network
``partitions''~\cite{spbz11a,spbz11b,spal11}.  Commutative Replicated Data
Types (CRDTs)~\cite{spbz11b}, for example, combine strong consistency with
immediate responses, in order to enable strong consistency even with
replication over wide-area networks.  \NAME exploits a similar,
workload-dependent property for achieving a different goal, which allows
\NAME to run on an untrusted server.

SPORC~\cite{fzff10} is a group collaboration system where operations do not
need to be executed in the same order at every client by virtue of
employing \emph{operational transforms}.  The latter concept allows for
shifting operations to a different position in an execution by transforming
them according to properties of the skipped operations.  Differently
ordered and transformed variants of a common sequence converge to the same
end state.
SPORC is stated to provide fork-* linearizability~\cite{limaz07}, which is
almost the same as weak \forklin~\cite{cakesh11}; both notions are strict
relaxations of \forklin that permit concurrent
operations to proceed without blocking, such that protocols become
wait-free.  The increased concurrency is traded for weaker consistency, as
up to one diverging operation may exist between two clients.  Moreover,
there is no formal analysis for SPORC.  As in BST, SPORC addresses only the
updates of client states and does \emph{not} consider local outputs;
however, for showing linearizability, one has to consider the responses of
operations.

The subsequent extension of SPORC to Frientegrity~\cite{fbfw12} leverages
read/write operations on storage objects to a complete social network that
may be hosted on an untrusted provider.  Frientegrity provides fork-*
linearizability as consistency condition like SPORC and adds many further
features beyond our interest.  However, the consistency and integrity
verification properties are the same as for SPORC.

FAUST~\cite{cakesh11} and Venus~\cite{scckms10}, mentioned before, never
block clients and enjoy eventual consistency, but guarantee only weak
\forklin.

In contrast to these protocols, \NAME ensures the stron\-ger \forklin
condition, where every operation is consistent as soon as it completes and
no client is ever in danger of acting upon receiving an arbitrary output.
In terms of expressiveness, SPORC is neither weaker nor stronger than
\NAME: On one hand, SPORC seems more general as it never blocks clients
even for operations that do not appear to commute.  On the other hand,
SPORC is limited to functions with transformable operations and mandates
that all operations are invertible.  Therefore SPORC cannot address
services with conflicting operations, which exist in many realistic service
specifications~\cite{cashsh07}.

VICOS~\cite{brcakn15} protects the integrity and consistency verification
of a generic cloud-object storage service.  It extends the protocol of this
work and shows how to apply it in a practical deployment.

\subsection{Organization and relation to previous version}
\label{ssec:organization}

A predecessor of this paper~\cite{cacohr14} considered only
\emph{commutative} operations instead of conflict-free ones.  Brandenburger
et al.~\cite{brcakn15} have subsequently introduced \emph{conflict-free}
operations with VICOS and shown that it is sufficient to abort only when a
conflict occurs but not for all commuting operations.  We take this up here
because the notion is more general than commutativity for consistency
verification.

Furthermore, the \emph{authenticated} version of \NAME in~\cite{cacohr14},
which shifted the state from the clients to the server and appeared in the
earlier version, is not contained here.  The reason lies in the lack of
formalization, which would go beyond the scope of this version.
VICOS~\cite{brcakn15} also exploits authenticated data types for keeping
the state remotely but does not present a formal consistency analysis.  The
focus of this work is on a proof for achieving \forklin with \NAME.

This paper continues by first introducing the notation and basic concepts
in Section~\ref{sec:def}.  The subsequent section pre\-sents \NAME and
discusses its properties.  A detailed analysis of \NAME 
follows in Section~\ref{sec:analysis}.
Finally Section~\ref{sec:conclusion} concludes the paper.

\section{Definitions}
\label{sec:def}

\subsection{System model}
We consider an asynchronous distributed system with $n$ clients, $C_1,
\dots, C_n$ and a server~$S$, modeled as processes.  Each client is
connected to the server through an asynchronous, reliable
communication channel that respects FIFO order.  A protocol specifies
the operations of the processes.  All clients are \emph{correct} and
follow the protocol, whereas $S$ operates in one of two modes: either
she is \emph{correct} and follows the protocol or she is
\emph{Byzantine} and may deviate arbitrarily from the specification.

\subsection{Functionality}
We consider a deterministic \emph{functionality}~$F$ (also called a type)
defined over a set of
\emph{states}~\CS and a set of \emph{operations}~\CO. $F$ takes as
arguments a state $s \in \CS$ and an operation~$o \in \CO$ and returns
a tuple $(s', r)$, where $s' \in \CS$ is a state that reflects any
changes that $o$ caused to $s$ and $r \in \CR$ is a response to $o$
\[
(s', r) \becomes F(s, o).
\]
This is also called the \emph{sequential specification} of~$F$.

We extend this notation for executing a sequence of operations
$\langle o_1, \dots, o_k \rangle$, starting from an initial
state~$s_0$, and write
\[
  (s', r) \ = \ F(s_0, \langle o_1, \dots, o_k \rangle)
\]
for $(s_i, r_i) = F(s_{i-1}, o_i)$ with $i = 1, \dots, k$ and $(s', r)
= (s_k, r_k)$.  Note that an operation in \CO may represent a batch
of multiple application-level operations.

\subsection{Operations and conflicts}

Conflicts between operations of~$F$ play an important role in protocols
that may execute multiple operations concurrently and have been studied
intensively in the context of multi-version concurrency control for
databases~\cite{WeikumV02} as well as in concurrent and distributed
computing~\cite{HerlihyS08}.  In this work, an operation $o_1 \in \CO$ is
said to \emph{conflict} with an operation $o_2 \in \CO$ \emph{in a state~$s
  \in \CS$} if and only if the presence of $o_1$ before $o_2$ influences
the return value of~$o_2$.
In other words, if $C_1$ executes $o_1$ and $o_1$ does not
conflict with $o_2$ executed by $C_2$, then $C_2$ can go ahead and generate
output for $o_2$ without waiting until $o_1$ finishes.  This improves the
throughput of \NAME compared to earlier protocols.

Conflicts are asymmetric.  Formally, $o_1$ does \emph{not conflict} with
$o_2$ \emph{in a state~$s$} if and only if, for
\begin{align*}
  (s', r_1)  \ &\gets \ F(s, o_1) \\
  (s'', r_2) \ &\gets \ F(s', o_2) \\
  (t, q)     \ &\gets \ F(s, o_2)
\end{align*}
it holds
\[
  r_2 = q.
\]
Furthermore, we say that $o_1$ does \emph{not conflict} with~$o_2$ whenever
$o_1$ does not conflict with $o_2$ in any state of~$F$.
Commuting operations (as considered in earlier work) never conflict, but
two non-conflicting operations may not commute.

Not only individual operations, also sequences of them may be conflict-free
to each other in this sense.  Suppose two sequences~$\mu$ and $\rho$
consisting of operations in \CO are mixed together into one sequence~$\pi$
such that the partial order among the operations from $\mu$ and from
$\rho$ is retained in~$\pi$, respectively.  If executing~$\pi$ starting
from a state~$s$ gives the same responses for all operations of $\rho$ as
in every other such mixed sequence, in particular $\mu \concat \rho$
and $\rho \concat \mu$, where $\concat$ denotes concatenation, we say
that $\mu$ does \emph{not conflict} with $\rho$ \emph{in state~$s$}.
Analogously, we say that $\mu$ does \emph{not conflict} with $\rho$ if
$\mu$ does not conflict with $\rho$ in any state.

We define a Boolean predicate $\op{conflict}_F(s, \mu, \rho)$ that
returns \true if and only if a sequence of operations~$\mu$ conflicts
with a sequence~$\rho$ in~$s$ according to~$F$.  W.l.o.g. we assume that
all operations of~$F$ and the predicate $\op{conflict}_F$ are efficiently
computable.

Observe that changes to the state of $F$ are not considered in the conflict
relation.  In particular, two operation sequences $\mu$ and $\rho$ that
``write'' to the same low-level value do not conflict as long as the
responses of $\rho$ remain the same as in the absence of~$\mu$.  This might
look different from the usual notion of conflicts considered in other
works, because the effects on the underlying state are considered
implicitly, only through operations of~$\rho$ whose output depends on the
state.  (Indeed, if $\mu$ modifies the state and an operation in $\rho$
returns the complete state, then $\mu$ always conflicts with $\rho$.)  As
will become clear later, the conflict relation introduced here is adequate
for consistency verification.

\subsection{Abortable services}
When operations of $F$ conflict, a protocol may either decide to block
or to abort.  Aborting and giving the client a chance to retry the
operation at his own rate often has advantages compared to blocking,
which might delay an application in unexpected ways.  

As in previous work that permitted aborts~\cite{afhht07,mdss09}, we allow
operations to abort and augment~$F: \CS \times \CO \to \CS \times \CR$ to
an \emph{abortable} functionality~$G$ accordingly.  $G: \CS \times \CO \to
\CS \times \CR_\bot$ is defined over the same set of states~\CS and
operations~\CO as $F$, but returns a tuple defined over \CS and response
set $\CR_\bot = \CR \cup \{\bot\}$.  $G$ will usually return the same
output as~$F$, but it may also return $\bot$ and leave the state unchanged,
denoting that a client is not able to execute~$F$.  Hence, $G$ is a
relation and satisfies
\[
  G(s, o) = \bigl\{ (s, \bot) , \, F(s, o) \bigr\}.
\]

The abortable~$G$ inherits most properties of~$F$ apart from its
deterministic specification.  In particular, since $G$ is not
deterministic, a sequence of operations no longer uniquely determines the
resulting state and response value.  Whenever the abortable~$G$ is used in
a protocol, one has to explicitly require that if $G$ is accessed only
sequentially, then $G$ should never abort, i.e., it should always behave
like~$F$.

In the sequel, when we refer to a generic \emph{functionality} $\Lambda$,
this may represent the deterministic~$F$ or its abortable extension~$G$.

Abortable functionalities are related to obstruction-free
objects~\cite{afhht07,helumo03} in shared-memory systems subject to
concurrent operations; such objects also guarantee that every client
operation completes assuming the client eventually runs in isolation.

\subsection{Operations and histories}
\label{sec:defops}

The clients interact with a functionality $\Lambda$ through
\emph{operations} provided by $\Lambda$.  As operations take time, they are
represented by two events occurring at the client, an \emph{invocation} and
a \emph{response}.  A \emph{history} of an execution~$\sigma$ consists of
the sequence of invocations and responses of $\Lambda$ occurring in
$\sigma$.  An operation is \emph{complete} in a history if it has a
matching response.

An operation~$o$ \emph{precedes} another operation~$o'$ in a sequence
of events $\sigma$, denoted $o <_\sigma o'$, whenever $o$ completes
before $o'$ is invoked in $\sigma$. A sequence of events $\pi$
\emph{preserves the real-time order} of a history~$\sigma$ if for
every two operations $o$ and $o'$ in $\pi$, if $o <_\sigma o'$ then
$o<_\pi o'$.  Two operations are \emph{concurrent} if neither one of
them precedes the other.  A sequence of events is \emph{sequential} if
it does not contain concurrent operations.  For a sequence of
events~$\sigma$, the subsequence of $\sigma$ consisting only of events
occurring at client~$C_i$ is denoted by $\sigma|_{C_i}$ (we use the
symbol $|$ as a projection operator).  For some operation
$\textit{o}$, the prefix of $\sigma$ that ends with the last event of
$o$ is denoted by $\sigma|^{o}$.

An operation $o$ is said to be \emph{contained in} a sequence of
events~$\sigma$, denoted $o \in \sigma$, whenever at least one event
of $o$ is in $\sigma$.  We often simplify the terminology by
exploiting that every \emph{sequential} sequence of events corresponds
naturally to a sequence of operations, and that analogously every
sequence of operations corresponds to a sequential sequence of events.

An execution is \emph{well-formed} if the events at each client are
alternating invocations and matching responses, starting with an
invocation.  An execution is \emph{fair}, informally, if it does not
halt prematurely when there are still steps to be taken or messages to
be delivered (see the standard literature for a formal
definition~\cite{Lynch96}).  We are interested in a protocol where the
clients never block, though some operations may be aborted and thus
will not complete regularly.  We call a protocol \emph{wait-free} if
in every history where the server is correct, every operation by any
client completes~\cite{herlih91}.

\subsection{Consistency properties}
\label{sec:defconsistency}

We use the standard notion of \emph{linearizability}~\cite{herwin90}, which
requires that the operations of all clients appear to execute atomically in
one sequence.  \emph{Fork-linearizability}~\cite{mazsha02,cashsh07} relaxes
the condition of one sequence and extends it to permit multiple ``forks''
of an execution.  Under \forklin, every client observes a linearizable
history and when an operation is observed by multiple clients, the
operation sequence occurring before that operation is the same.  In other
words, the history of operations forms a tree whose branches are the forks,
and the operations on the path from the root to every leaf are
linearizable, and every client observes exactly the operations on the path
to one leaf.  For simplicity we leave out incomplete operations in
(fork-)linearizability; although they could readily be included as in other
works, we feel this would overly complicate the analysis of the protocol.

\begin{definition}[View]\label{def:view}
  A sequence of events $\pi$ is called a \emph{view} of a history
  $\sigma$ at a client $C_i$ w.r.t. a functionality $\Lambda$ if:
\begin{enumerate}\shortlist
\item $\pi$ is a sequential permutation of some subsequence
of complete operations in $\sigma$;
\item all complete operations executed by $C_i$ appear in $\pi$; and
\item $\pi$ satisfies the sequential specification of $\Lambda$.
\end{enumerate}
\end{definition}

\begin{definition}[Linearizability~\cite{herwin90}]\label{def:lin}
  A history $\sigma$ is \emph{linearizable w.r.t. a functionality
    $\Lambda$} if there exists a sequence of events $\pi$ such that:
\begin{enumerate}\shortlist
\item $\pi$ is a view of $\sigma$ at all clients w.r.t. $\Lambda$; and
\item $\pi$ preserves the real-time order of $\sigma$.
\end{enumerate}
\end{definition}

\begin{definition}[Fork-linearizability~\cite{mazsha02}]\label{def:forklin}
  A history $\sigma$ is \emph{fork-linearizable w.r.t. a
    functionality~$\Lambda$} if for each client $C_i$ there exists a
  sequence of events $\pi_i$ such that:
\begin{enumerate}\shortlist
  \item $\pi_i$ is a view of $\sigma$ at $C_i$ w.r.t. $\Lambda$;
  \item $\pi_i$ preserves real-time order of $\sigma$; and
  \item for every client $C_j$ and every operation $o \in \pi_i \cap \pi_j$ 
  it holds that $\pi_i|^o = \pi_j|^o$.
\end{enumerate}
\end{definition}

Finally, we recall the concept of a \textit{fork-linearizable Byzantine
  emulation}~\cite{cashsh07}.  It summarizes the requirements put on our
protocols, which runs between the clients and an untrusted server.  This
notion means that when the server is correct, the service should guarantee
the standard notion of linearizability; otherwise, it should ensure
\forklin.

\begin{definition}[Fork-linearizable Byzantine emulation~\cite{cashsh07}]
  We say that a protocol~$P$ for a set of clients \emph{emulates} a
  functionality~$\Lambda$ \emph{on a Byzantine server~$S$ with
    \forklin} if:
  \begin{enumerate}\shortlist
  \item in every fair and well-formed execution of $P$, the sequence of
    events observed by the clients is fork-linearizable with respect
    to~$\Lambda$; and
  \item if $S$ is correct, then the execution is linearizable w.r.t.~$\Lambda$.
  \end{enumerate}
\end{definition}

\subsection{Cryptographic primitives}
\label{sec:defcrypto}

As the focus of this work is on concurrency and correctness and not on
cryptography, we model \emph{hash functions} and \emph{digital
  signature schemes} as ideal, deterministic functionalities
implemented by a distributed oracle.

A \emph{hash function} maps a bit string of arbitrary length to a
short, unique representation.  The functionality provides only a
single operation \op{hash}; its invocation takes a bit string $x$ as
parameter and returns an integer $h$ with the response.  The
implementation maintains a list $L$ of all $x$ that have been queried
so far.  When the invocation contains $x\in L$, then \op{hash}
responds with the index of $x$ in $L$; otherwise, \op{hash} appends
$x$ to $L$ and returns its index.  This ideal implementation models
only collision resistance but no other properties of real hash
functions.

The functionality of the \emph{digital signature scheme} provides two
operations, $\op{sign}_i$ and $\op{verify}_i$.  The invocation of
$\op{sign}_i$ specifies the index $i$ of a client and takes a bit string~$m
\in \{0,1\}^*$ as input and returns a signature $\sigma \in \{0,1\}^*$ with
the response.  Only $C_i$ may invoke $\op{sign}_i$.  The operation
$\op{verify}_i$ takes a putative signature~$\sigma$ and a bit string~$m$ as
parameters and returns a Boolean value with the response.  Its
implementation satisfies that $\op{verify}_i(\sigma ,m)$ returns \true for
any $i \in \{1,\dots,n\}$ and $m \in \{0,1\}^*$ if and only if $C_i$ has
executed $\op{sign}_i(m)$ and obtained $\sigma$ before; otherwise,
$\op{verify}_i(\sigma ,m)$ returns \false.  Every client as well as $S$ may
invoke \op{verify}.  The signature scheme may be implemented analogously to
the hash function.

\section{The conflict-free operation verification protocol}
\label{sec:protocol}

\subsection{Protocol description}
\label{ssec:protocol}

\para{Notation}
The function $\op{length}(L)$ for a list~$L$ denotes the number of elements
in~$L$ and $\|$ stands for the concatenation of strings.  Several variables
are \emph{dynamic arrays} or \emph{maps}, which associate keys to values.
A value~$v$ is stored in a map $H$ by assigning it to a key~$k$, denoted
$H[k] \becomes v$; if no value has been assigned to a key, the map
returns~$\bot$.  For simplicity, $\bot$ also stands for the empty bit
string.
Recall that $G$ is the abortable extension of functionality~$F$.

\para{Overview}
The pseudocode of \NAME for the clients and the server is presented in
Algorithms~\ref{alg:client}--\ref{alg:server}.  We assume that the
execution of each client is well-formed and fair.

\NAME adopts the structure of previous protocols that guarantee
fork-linearizable semantics~\cite{mazsha02,wisish09,cachin11}.  It aims at
obtaining a globally consistent order for the operations of all clients, as
determined by the server.  Every client maintains a copy of the service
state and executes all operations locally.

When a client~$C_i$ \emph{invokes} an operation~$o$, he sends an
$\str{invoke}$ message to the server~$S$
(L\ref{c:op:begin}--L\ref{c:op:end}).  He expects to receive a
$\str{reply}$ message from~$S$ telling him about the position of $o$ in the
global sequence of operations.  The message contains the operations that
are \emph{pending} for~$o$, that is, operations which go beyond the prefix
of the history that $C_i$ has already verified for consistency.  These
pending operations are ordered before~$o$ by a correct~$S$, but $C_i$ may
not yet know about some of them.  (A Byzantine~$S$ may introduce
consistency violations here.)  We distinguish between \emph{pending-other}
operations invoked by other clients and \emph{pending-self} operations,
which are operations executed by $C_i$ up to~$o$.

When $C_i$ receives the \str{reply} message with~$o$, he verifies whether
the data from the server is consistent and, if everything is valid, he
commits~$o$.  \NAME uses \textbf{assert} statements for verification.  If
any of these steps fail, the formal protocol simply halts; in practice, the
clients would then recover the service state, abandon the faulty~$S$, and
switch to another provider.
In order to ensure \forklin for the response values, the client first
executes $o$ by \emph{simulating} the pending-self operations and $o$
according to $F$ (that is, without updating the locally held state).  If
the pending-other operations do \emph{not conflict} with the pending-self
operations and $o$, then he declares $o$ to be \emph{successful} and
outputs the response~$r$ according to~$F$, as resulting from the simulated
operations.  Otherwise, the client \emph{aborts} $o$ and the response is~$r
= \bot$.  According to this, the \emph{status} of $o$ is a value in $\CZ =
\{ \str{success}, \str{abort} \}$.  Through these steps the client
\emph{commits}~$o$.  Then he sends a corresponding $\str{commit}$ message
to~$S$ and outputs~$r$.

The (correct) server records the committed operation and relays it to all
clients via a $\str{broadcast}$ message
(L\ref{s:invoke:begin}--L\ref{s:invoke:end}).  When the client receives
such a broadcast operation, he verifies that it is consistent with
everything the server told him so far. If this verification succeeds, we
say that the client \emph{confirms} the operation.  If the operation's
status was \str{success}, then the client executes it and \emph{applies} it
to his local state (L\ref{c:bc:begin}--L\ref{c:bc:end}).

\para{Data structures}
Every client locally maintains a set of variables during the protocol
(L\ref{c:state:begin}--L\ref{c:state:end}).  The state $s\in\CS$ is the
result of applying all confirmed and successful operations, received in
\str{broadcast} messages, to the initial state~$s_0$.  Variable~$c$ stores
the sequence number of the last operation that the client has confirmed.
$H$ is a map containing a \emph{hash chain} computed over the operation
sequence as announced by~$S$ to~$C_i$.  The contents of $H$ are indexed by
the sequence number of the operations.  Entry $H[l]$ is computed as
$\op{hash}(H[l-1] \| o \| l \| i)$, with $H[0] = \nil$, and represents an
operation~$o$ with sequence number~$l$ executed by~$C_i$.  The hash chain
allows for fast comparisons between the histories of two clients: if they
obtain the same hash-chain value, then both have confirmed the same
sequence of operations.

The client sets a variable~$\bar{o}$ to~$o$ whenever it has invoked an
operation $o$ but not yet completed it; at other times $\bar{o}$
is~$\bot$.  Variable~$Z$ maps the sequence number of every operation that
the client has executed himself to its status.  The client only needs the
entries in~$Z$ with indices greater than~$c$.

The (correct) server also keeps several variables locally
(L\ref{s:state:begin}--L\ref{s:state:end}).  Variable~$t$ determines the
global sequence number for the invoked operations and $b$ denotes the
sequence number of the last broadcast operation.  The latter ensures that
$S$ disseminates operations to clients in the global order.  Furthermore,
she stores the invoked operations in a map~$I$ and the completed operations
in a map~$O$, both indexed by sequence number.

\begin{algo*}
\vbox{
\small
\begin{numbertabbing}\reset
xxxx\=xxxx\=xxxx\=xxxx\=xxxx\=xxxx\=xxxx\kill
\textbf{State} \label{c:state:begin}\\
\> $\bar{o} \in \CO \cup \{\bot\}$: the operation being executed currently 
   or~$\bot$ if no operation runs, initially~$\bot$ \label{}\\
\> $c \in \N_0$: sequence number of the last operation that has been confirmed, 
   initially~0 \label{}\\
\> $H: \N_0 \to \{0,1\}^*$: 
   hash chain (see text), initially containing only $H[0] = \nil$ \label{}\\
\> $Z: \N_0 \to \CZ  \cup \{\bot\}$: 
   status map (see text), initially empty  \label{}\\
\> $s \in \CS$: current state, after applying operations, initially $s_0$ \label{c:state:end}\\
\\
\textbf{upon invocation} $o$ \textbf{do}
   \` // \emph{invoke} operation $o$ \label{c:op:begin}\\
\> $\bar{o} \becomes o$ \label{}\\
\> $\tau \becomes \op{sign}_i(\str{invoke} \| o \| i)$ \label{c:signinvoke}\\
\> send message \msg{invoke}{o, \tau} to $S$ \label{c:op:end}\\
\\
\textbf{upon} receiving message \msg{reply}{\var{Pend}} from $S$ \textbf{do} 
   \` // the last operation in \var{Pend} should be $\bar{o}$ 
   \label{c:reply:begin}\\
\> $\var{Pend-other} \becomes \langle \rangle$ 
   \` // list of pending-other operations \label{}\\
\> $\var{Pend-self} \becomes \langle \rangle$ 
   \` // list of successful pending-self operations \label{}\\
\> $k \becomes 1$ \label{}\\ 
\> \textbf{while} $k \leq \op{length}(\var{Pend})$ \textbf{do} 
   \label{c:loop:begin}\\
\> \> $(o, j, \tau) \becomes \var{Pend}[k]$ \label{}\\
\> \> $l \becomes c+k$ 
      \` // promised sequence number of $o$ \label{c:index}\\
\> \> \textbf{assert} $\op{verify}_j(\tau, \str{invoke} \| o \| j)$ 
      \label{c:verifyinvoke}\\
\> \> \textbf{if} $H[l] = \bot$ \textbf{then} \label{c:chain:1a}\\
\> \> \> $H[l] \becomes \op{hash}(H[l-1] \| o \| l \| j)$ 
         \` // extend hash chain \label{c:chain:1b}\\
\> \> \textbf{else} \label{}\\
\> \> \> \textbf{assert} $H[l] = \op{hash}(H[l-1] \| o \| l \| j)$
         \` // server replies must be consistent \label{c:chain:1test}\\
\> \> \textbf{if} $j = i \land k < \op{length}(\var{Pend})
                   \land Z[l] = \str{success} $ \textbf{then}
                   \label{c:pend:begin}\\ 
\> \> \> $\var{Pend-self} \becomes \var{Pend-self} \concat \langle o \rangle$ \label{c:pend:self}\\
\> \> \textbf{else if} $j \neq i$ \textbf{then} \label{}\\
\> \> \> $\var{Pend-other} \becomes  \var{Pend-other} \concat \langle o \rangle$ \label{c:pend:end}\\
\> \> $k \becomes k + 1$ \label{c:loop:end}\\
\> // variables $o$, $j$, and $l = c + \op{length}(\var{Pend})$ 
         keep their values \label{}\\
\> \textbf{assert} $k > 1 \land o = \bar{o} \land j = i$
      \` // last pending operation must equal the current operation 
      \label{c:otest}\\
\> \textbf{if} \textbf{not} $\op{conflict}_F(s, \var{Pend-other},  
      \var{Pend-self} \concat \langle o \rangle)$ 
   \textbf{then} 
   \` // $o = \bar{o}$ is the current operation \label{c:conflict:no}\\
\> \> $(s', r) \becomes F(s, \var{Pend-self} \concat \langle o \rangle)$
   \` // compute response to $o$ and ignore resulting state \label{c:simulate}\\
\> \>  $Z[l] \becomes \str{success}$ \label{c:success}\\
\> \textbf{else} \label{c:conflict:yes}\\
\> \> $r \becomes \bot$ \label{}\\
\> \> $Z[l] \becomes \str{abort}$ \label{c:conflict:end}\\
\> $\phi \becomes 
   \op{sign}_i\bigl(\str{commit} \| o \| l \| H[l] \| Z[l]\bigr)$ 
   \` // \emph{commit} operation~$\bar{o}$ \label{c:commit}\\
\> send message \msg{commit}{o, l, H[l], Z[l], \phi} to $S$ \label{}\\
\> $\bar{o} \becomes \bot $ \label{}\\
\> \textbf{return} $r$ 
   \` // \emph{complete} operation~$\bar{o}$ \label{c:reply:end}\\
\\
\textbf{upon} receiving message \msg{broadcast}{o, l, h, z, \phi, j} from $S$ \textbf{do} \label{c:bc:begin}\\
\> \textbf{assert} $l = c+1 \land
              \op{verify}_j(\phi, \str{commit} \| o \| l \| h \| z)$
   \` // start to \emph{confirm} operation $o$
   \label{c:confirm}\\
\> \textbf{if} $H[l] = \bot$ \textbf{then} 
   \` // operation $o$ has never been pending at $C_i$ \label{c:chain:2a}\\
\> \> $H[l] \becomes \op{hash}(H[l-1] \| o \| l \| j)$ \label{c:chain:2b}\\
\> \textbf{assert} $h = H[l]$
   \` // if this holds, then $o$ is \emph{confirmed} \label{c:chain:2test}\\
\> \textbf{if} $z = \str{success}$  \textbf{then} 
      \` // apply $o$ only if successful \label{c:ifapply}\\
\> \> $(s, r') \becomes F(s, o)$
   \` // \emph{apply} operation $o$ and ignore response \label{c:doapply}\\
\> $c \becomes c + 1$ \label{c:bc:end}
\end{numbertabbing}
}
\caption{Conflict-free operation verification protocol (client~$C_i$)}
\label{alg:client}
\end{algo*}

\begin{algo*}
\vbox{
\small
\begin{numbertabbing}\resetto{100}
xxxx\=xxxx\=xxxx\=xxxx\=xxxx\=xxxx\=xxxx\kill
\textbf{State} \label{s:state:begin}\\
\> $t \in \N_0$: sequence number of the last invoked operation, 
   initially 0 \label{}\\
\> $b \in \N_0$: sequence number of the last broadcast operation, 
   initially 0 \label{}\\
\> $I: \N \to \CO \times \N_0 \times \{0,1\}^*$: 
   invoked operations (see text), initially empty \label{}\\
\> $O: \N \to \CO \times \{0,1\}^* \times \CZ \times \{0,1\}^* \times \N$: 
   committed operations (see text), initially empty \label{s:state:end}\\
\\
\textbf{upon} receiving message $\msg{invoke}{o, \tau}$ from $C_i$ \textbf{do} \label{s:invoke:begin}\\
\> $t\becomes t +1$ \label{}\\
\> $I[t] \becomes (o, i, \tau)$ \label{}\\
\> $\var{Pend} \becomes \langle I[b+1], \ldots, I[t] \rangle$ 
   \` // include non-committed operations and~$o$ \label{}\\
\> send message \msg{reply}{\var{Pend}} to $C_i$ \label{s:invoke:end}\\
\\
\textbf{upon} receiving message $\msg{commit}{o, l, h, z, \phi}$ from $C_i$ \textbf{do} \label{}\\
\> $O[l] \becomes (o, h, z, \phi, i)$ \label{s:store}\\
\> \textbf{while} $O[b+1] \neq \bot$ \textbf{do} 
   \` // broadcast operations ordered by their sequence number
   \label{s:loop:begin}\\
\> \> $b \becomes b + 1$ \label{}\\
\> \> $(o', h', z', \phi', j) \becomes O[b]$ \label{}\\
\> \> send message \msg{broadcast}{o', b, h', z', \phi', j} to all clients
      \label{s:loop:end}
\end{numbertabbing}
}
\caption{Conflict-free operation verification protocol (server~$S$)}
\label{alg:server}
\end{algo*}

\para{Protocol}
When client~$C_i$ invokes an operation~$o$
(L\ref{c:op:begin}--L\ref{c:op:end}), he stores it in~$\bar{o}$ and sends
an \str{invoke} message to $S$ containing~$o$ and~$\tau$, a digital
signature computed over $o$ and~$i$.  In turn, a correct $S$ sends a
\str{reply} message with the list~$\var{Pend}$ of pending operations
(L\ref{s:invoke:begin}--L\ref{s:invoke:end}); the operations have sequence
numbers~$c+1$, $c+2$, \dots.  Upon receiving a \str{reply} message, the
client checks that $\var{Pend}$ is consistent with any previously sent
operations and uses $\var{Pend}$ to assemble the pending-other
operations~$\var{Pend-other}$ and the successful pending-self
operations~$\var{Pend-self}$.  He then determines whether $o$ can be
executed or has to be aborted (L\ref{c:reply:begin}--L\ref{c:reply:end}).

In particular, during the loop in Algorithm~\ref{alg:client}
(L\ref{c:loop:begin}--L\ref{c:loop:end}), for every operation~$o$
in~$\var{Pend}$, client~$C_i$ determines its sequence number~$l$ and
verifies from the \str{invoke} signature that~$o$ was indeed invoked by
$C_j$ (L\ref{c:index}--L\ref{c:verifyinvoke}).  He computes the entry
of~$o$ in the hash chain from $o$, $l$, $j$, and $H[l-1]$.  If $H[l] =
\bot$, then $C_i$ stores the hash value there.  Otherwise, $H[l]$ has
already been set and $C_i$ verifies that the hash values are equal; this
means that $o$ is consistent with the pending operation(s) that $S$ has
sent previously with indices up to~$l$
(L\ref{c:chain:1a}--L\ref{c:chain:1test}).

If operation~$o$ is his own and its saved status in $Z[l]$ was
\str{success}, then he appends it to~$\var{Pend-self}$
(L\ref{c:pend:begin}--L\ref{c:pend:self}). The client remembers the status
of his own operations in $Z$, since $\op{conflict}_F$ depends on the state
and that could have changed if he applied operations after committing~$o$.
Operations of other clients from \var{Pend} are added to \var{Pend-other}
(L\ref{c:pend:end}).

Finally, when $C_i$ reaches the end of~$\var{Pend}$, he checks that
$\var{Pend}$ is not empty and that it contains $o = \bar{o}$ at the last
position (L\ref{c:otest}).  He then tests whether the pending-other
operations~$\var{Pend-other}$ do not conflict with $\var{Pend-self} \concat
\langle o \rangle$ in state~$s$, his state resulting from the confirmed
operations (L\ref{c:conflict:no}). If there is no conflict, he records the
status of $o$ as \str{success} in $Z[l]$ and computes the response~$r$ by
executing $\var{Pend-self} \concat \langle o \rangle$ starting from~$s$
(L\ref{c:simulate}--L\ref{c:success}).  Otherwise, if $\var{Pend-other}$
conflicts with $o$, he records the status of $o$ as $Z[l] \becomes
\str{abort}$ and sets $r \becomes \bot$
(L\ref{c:conflict:yes}--L\ref{c:conflict:end}).  Then $C_i$ signs $o$
together with its sequence number, status, and hash chain entry~$H[l]$,
includes all values in the \str{commit} message sent to~$S$, and
returns~$r$ (L\ref{c:commit}--L\ref{c:reply:end}).  Through these steps
$C_i$ \emph{commits}~$o$.

Upon receiving a \str{commit} message for an operation $o$ with sequence
number~$l$, the (correct) server records its content as~$O[l]$ in the map
of committed operations (L\ref{s:store}).  Then she is supposed to send a
\str{broadcast} message containing $O[l]$ to the clients.  She waits with
this until she has received \str{commit} messages for all operations with
sequence number less than $l$ and has also broadcast them
(L\ref{s:loop:begin}--L\ref{s:loop:end}).  This ensures that completed
operations are disseminated in the global order to all clients, exploiting
the FIFO channels between the correct $S$ and the clients.

In a \str{broadcast} message received by client~$C_i$ (L\ref{c:bc:begin}),
the committed operation is represented by a tuple $(o, l, h, z, \phi, j)$.
Client~$C_i$ conducts several verification steps.  If successful, we say
$o$ is \emph{confirmed}.  If $o$ did not abort, then $C_i$ subsequently
\emph{applies}~$o$ to his state~$s$.  In more detail, the client first
verifies that the sequence number~$l$ is the next operation according
to~$c$ (L\ref{c:confirm}); hence, $o$ follows the global order and the
server did not omit any operations.  Second, he uses the \str{commit}
signature~$\phi$ in the message to verify that $C_j$ indeed committed~$o$
(L\ref{c:confirm}).  Lastly, $C_i$ computes his own hash-chain entry $H[l]$
for~$o$ and asserts that it is equal to the hash-chain value~$h$ from the
message (L\ref{c:chain:2a}--L\ref{c:chain:2test}).  This ensures that $C_i$
and $C_j$ have received consistent operations from $S$ up to $o$.  Once the
verification succeeds, the client applies~$o$ to his state~$s$ only if its
status~$z$ was \str{success}, that is, when $C_j$ has not aborted~$o$
(L\ref{c:ifapply}--L\ref{c:doapply}).

Observe that $C_i$ must output the response of a successful operation after
receiving the \str{reply} message (L\ref{c:reply:end}).  To satisfy
\forklin for the output, the view of $C_i$ must contain at least its
pending-self operations and the output value must not change even if a
faulty server would cause the other clients to commit the pending-other
operations differently than announced to~$C_i$.  The state ($s'$ in
L\ref{c:simulate}) computed by $C_i$ is ignored though, as it is computed
with the pending-other operations skipped (hence, it may not
reflect all operations in~$C_i$'s view).

\subsection{Features of \NAME}
\label{ssec:features}

\para{Conflicts in operation sequences}
Consider the following example $F$ of a counter restricted to non-negative
values: Its state consists of an integer~$s$; an $\op{add}(x)$ operation
adds~$x$ to $s$ and returns \true; a $\op{dec}(x)$ operation subtracts~$x$
from $s$ and returns \true if $x \leq s$, but does nothing and returns
\false if $x > s$.  

We use this to illustrate three properties of \NAME, where $S$ is always
correct in the examples.  Assume all client operations have completed and
that the state (after applying all operations) at $C_i$ is~$s=7$.
\begin{enumerate}\shortlist
\item Suppose $C_i$ executes $\op{add}(3)$ and the \str{reply} message
  contains a pending operation $\op{dec}(10)$.  The operation of $C_i$
  succeeds and is executed because no \op{add} or \op{dec} operation
  conflicts with $\op{add}(3)$, as its response is always \true.  However,
  the operations $\op{add}(3)$ and $\op{dec}(10)$ do not commute in state~7
  because the response from $\op{dec}(10)$ differs in the two possible
  orderings (the resulting states also differ, but this is not relevant for
  our notion of a conflict).  This shows that testing for
  \emph{conflict-free operations} permits more executions to succeed than
  checking only commuting operations.  Hence, \NAME aborts fewer executions
  than protocols aborting all non-commuting operations.

\item Client~$C_i$ executes $\op{dec}(5)$ and subsequently $\op{dec}(4)$,
  while $\op{add}(3)$ by another client is pending at both times.  Note
  that $C_i$ executes $\op{dec}(5)$ successfully but aborts $\op{dec}(4)$
  because $\op{add}(3)$ conflicts with $\langle \op{dec}(5), \op{dec}(4)
  \rangle$ in state~7.

  However, considering $C_i$'s operations individually, $\op{add}(3)$ does
  not conflict with $\op{dec}(5)$ nor with $\op{dec}(4)$ in state~7 because
  their return values are the same when $\op{add}(3)$ is omitted (although
  the resulting states differ).  This shows why the client considers the
  \emph{sequence} of all successful \emph{pending-self} operations when
  testing for a conflict with the current operation.

\item Suppose that $C_i$ executes $\op{dec}(5)$ and $S$ reports the pending
  sequence $\langle \op{dec}(2), \op{dec}(1) \rangle$.  Thus, $C_i$ aborts
  $\op{dec}(5)$.  Although, when considered individually from state~7,
  $\op{dec}(2)$ does not conflict with $\op{dec}(5)$ and $\op{dec}(1)$
  neither conflicts with $\op{dec}(5)$, their concatenation conflicts with
  $\op{dec}(5)$ and thus $C_i$ aborts.  This illustrates why \NAME checks
  for a conflict between the \emph{sequence} of \emph{pending-other}
  operations and the target operation.
\end{enumerate}
Neither of these three properties is present in previous protocols 
(as also discussed in Section~\ref{ssec:bst}).

\para{Memory requirements}
For saving space, the client may garbage-collect entries of $H$ and $Z$
with sequence numbers smaller than~$c$.  The server can also save space by
removing the entries in $I$ and $O$ for the operations that she has
broadcast.  However, if new clients are allowed to enter the protocol, the
server should keep all operations in $O$ and broadcast them to new clients
upon their arrival.

With the above optimizations the client has to keep in memory only the last
applied operation and the pending operations in~$H$ and the pending-self
operations in~$Z$. The same holds for the server: the maximum number of
entries stored in $I$ and $O$ is proportional to the number of pending
operations at any client.

\para{Complexity}
In terms of \emph{communication} cost, every operation executed by a client
requires him to perform one round\-trip to the server: send an \str{invoke}
message and receive a \str{reply}.  For every executed operation the server
sends a \str{broadcast} message to all clients.  Thus, when $\ell$
operations are executed overall, the protocol basically takes $O(\ell n)$
messages, although subsequent broadcasts to the same client could be
batched until the client invokes the next message~\cite{brcakn15}.  Clients
do not communicate with each other in the protocol.  However, as soon as
they do, they benefit from \forklin and can easily discover a forking
attack by comparing their hash chains.

Messages \str{invoke}, \str{commit}, and \str{broadcast} are independent of
the number of clients and contain only a description of one operation,
while the \str{reply} message contains the list $\var{Pend}$ of pending
operations.  If even one client is slow, then the length of $\var{Pend}$
for all other clients grows proportionally to the number of further
operations they are executing.  To reduce the size of \str{reply} messages,
the client can remember all pending operations received from~$S$, and $S$
can send every pending operation only once.

The total computational cost, on the other hand, is $O(\ell n)$ for
executing $\ell$ operations of~$F$, and this cannot be reduced as easily.
The reason lies in the maintenance of the hash chain at all clients, which
must be updated for every operation.  
Moreover, if a large number of pending operations are present during an
operation, the verification cost of the client increases proportionally.

\para{Aborts and wait-freedom}
Every client executing \NAME may proceed with an operation~$o$ for~$F$ as
long as no pending operations of other clients conflict with~$o$.  Observe
that the response to $o$ obtained by the client reflects all of his own
operations executed so far, even if he has not yet confirmed or applied
them to his state because operations of other clients have not yet
completed.  After successfully executing~$o$, the client outputs the
response directly while processing the \str{reply} message from~$S$.
However, when the pending operations of other clients conflict with~$o$,
the response would differ.  Thus, the client aborts~$o$ and outputs~$\bot$
according to~$G$.

Hence, for $F$ where no operations or operation sequences conflict \NAME is
wait-free; in particular, this holds when all of them commute.  For
arbitrary~$F$, however, no fork-linearizable Byzantine emulation can be
wait-free~\cite{cashsh07}.  \NAME avoids blocking via the augmented
functionality~$G$.  Clients complete every operation in the sense of $G$,
which includes aborts; therefore, \NAME is wait-free for~$G$.  In other
words, regardless of whether an operation aborts or not, the client may
proceed executing further operations.

To mitigate the risk of conflicts, the clients may employ a
synchronization mechanism such as a contention manager, scheduler, or
a simple random waiting strategy.  Such synchronization is common for
services with strong consistency demands.  If one considers also
clients that may crash (outside our formal model), then the client
group has to be adjusted dynamically or a single crashed client might
hold up progress of other clients forever.  Previous work on the topic
has explored how a group manager or a peer-to-peer protocol may
control a group membership protocol~\cite{lkms04,scckms10}; these
methods apply also to~\NAME.

\subsection{Comparison to Blind Stone Tablet (BST)}
\label{ssec:bst}
The BST protocol~\cite{wisish09} is a direct predecessor of \NAME but has
several shortcomings and does not achieve all claimed properties, as
explained now.

BST considers transactions on a database, coordinated by the remote server.
A client first simulates a transaction using his own copy, potentially
generating local output, then undoes this transaction on his copy, and
coordinates with the server for committing the transaction.  From the
server's response he determines if a transaction individually commutes with
every other, pending transaction that was reported by the server as invoked
by different clients.  If there is a conflict, the client ``rolls back the
external effects'' of the transaction and basically aborts; otherwise, he
``commits'' the transaction (but without changing his database copy) and
relays it via the server to other clients.  When a client receives such a
relayed transaction, he applies the transaction to his database copy. At
this high level BST is similar to \NAME.

However, when considering the details, several limitations of BST become
apparent: First, a client applies his own transactions only after all
pending transactions by other clients have been applied to his own database
copy.  This means that when the client executes a transaction~$T_B$,
updates induced by an earlier transaction~$T_A$ of his may not yet be
reflected in the database copy because they may be held back by earlier
transactions of other clients, which were pending during the execution
of~$T_A$, but have not yet been applied.  Thus, the client might execute
$T_B$ (in the simulation step) from a \emph{wrong} state, and this may
yield incorrect output for a linearizable execution.  Checking for the
absence of conflicts between $T_B$ and other transactions may also use such
a faulty state.  Alternatively, the protocol should block until the changes
from $T_A$ are applied to the database copy, but then the protocol is no
longer ``wait-free'' as stated~\cite{wisish09}.

Second, the BST client checks conflicts between his current transaction and
the incoming (pending) ones individually, considering each one alone but
not as the intended execution sequence.  Like the first limitation, this
implies that the client could violate consistency.  In particular, the
second and third example executions above, used for illustrating the
conflicts in Section~\ref{ssec:features}, will fail and produce wrong
outputs in BST.

Third, the notion of ``trace consistency'' in the analysis of BST considers
only the \emph{database state} and the transactions that have been
\emph{executed} on the local state~\cite{wisish09}.  However to satisfy
\forklin one must consider the responses output by the client.  The formal
notion of linearizability does not even consider the state of a
functionality, only the views of the clients are relevant.  A transaction
may be applied long after the client received the response and acted on it.
Hence, proof sketch available for BST~\cite[Sec.~5.2]{wisish09} does not
establish \forklin.

\NAME extends BST and allows one client to execute multiple operations
without waiting, i.e., independently of the speed of other clients, as long
as the \emph{sequence} of pending operations by other clients jointly does
not conflict with the client's operations, considering the current service
state.  Moreover, the analysis of \NAME shows it is fork-linearizable for
all \emph{responses} output by clients.

\section{Analysis}
\label{sec:analysis}

This section establishes the key properties of conflict-free operation
verification protocol (\NAME) in
Algorithms~\ref{alg:client}--\ref{alg:server}.

The first theorem addresses executions with a correct server and its proof
appears in the next section.  For stating the this result, we define the
following notion of overlapping operations.  It is a refinement of a
sequential execution that additionally takes into account the event that a
client \emph{applies} one of \emph{its own} operations
(L\ref{c:bc:begin}--L\ref{c:bc:end}); we introduce the term that the
operation is \emph{self-applied} to denote the event that this occurs.
We say that two operations $o$ and $o'$ in a history~$\sigma$
\emph{overlap} whenever the invocation of $o$ occurs after $o'$ is invoked
and before $o'$ is self-applied in $\sigma$, or vice versa, the invocation
of $o'$ occurs after the invocation of $o$ is invoked and before $o$ is
self-applied.  An execution~$\sigma$ \emph{without overlaps} is one in
which no two operations overlap.
\begin{theorem}\label{thm:lin}
  If the server is correct, then the history of every execution of \NAME is
  linearizable w.r.t. the abortable functionality~$G$.  Furthermore, if the
  clients execute all operations without overlaps, then all histories of
  \NAME are linearizable w.r.t.~$F$ and no operations abort.
\end{theorem}

The second theorem addresses executions with a Byzantine server and
captures the key goal of \NAME.
\begin{theorem}\label{thm:forklin}
  In every well-formed execution of \NAME, the history of events
  observed by the clients is \forklin w.r.t. the abortable functionality~$G$.
\end{theorem}

Together these results imply our main result.  Recall that a Byzantine
emulation implies that the execution is linearizable when $S$ is correct
and that it is \forklin otherwise.
\begin{corollary}\label{cor:byzemul}
  \NAME emulates the abortable functionality~$G$ on a Byzantine server with
  \forklin.
\end{corollary}

In the analysis we use the following terminology.  When a client issues a
\str{commit} signature for some operation~$o$, we say that he
\emph{commits}~$o$.  The client's sequence number included in the signature
thus becomes the \emph{sequence number of~$o$}; note that with a
faulty~$S$, two different operations may be committed with the same
sequence number by separate clients.

\subsection{Operating with a correct server}
\label{sec:lin}

This section contains a proof for Theorem~\ref{thm:lin}, which assumes $S$
is correct.  In particular, we show that the output of every client
satisfies~$G$ also in executions with concurrent or overlapping operations.
The check for conflicts, applied after simulating the client's pending-self
operations, ensures that the client's response remains unchanged regardless
of whether the pending-other operations execute before the operation itself
or not.

\begin{lemma}\label{lem:correct}
  If the server is correct, then every history~$\sigma$ is linearizable
  w.r.t.~$G$.
\end{lemma}

\begin{proof}
  Recall that~$\sigma$ consists of invocation and response events.  We now
  explain how to construct a sequential permutation~$\pi$ of $\sigma$.  We
  often rely on the correspondence between the pair of invocation and
  response events of one operation in $\sigma$ and the operation itself;
  hence, we sometimes treat $\pi$ as a sequence of operations to simplify
  the terminology.

  For the construction of $\pi$, note that a client sends an $\str{invoke}$
  message with his operation~$o$ to the server (L\ref{c:op:end}), the
  server assigns a sequence number to $o$, and sends it back
  (L\ref{s:invoke:begin}--L\ref{s:invoke:end}).  Since $S$ is correct, this
  is also the sequence number of~$o$.  The client then computes the
  response and sends a signed \str{commit} message to~$S$, containing the
  operation and its sequence number, and also outputs the response
  (L\ref{c:reply:begin}--L\ref{c:reply:end}).  Let $\pi$ consist of all
  events in $\sigma$, ordered first by the sequence number of the
  corresponding operation and including the invocation before the response
  with the same sequence number.

  As the server is correct, she processes $\str{invoke}$ messages in the
  order they are received and assigns sequence numbers accordingly.  This
  implies that if an operation~$o'$ is invoked after an operation~$o$
  completes, then the sequence number of $o'$ is higher than~$o$'s.
  Hence, $\pi$ preserves the real-time order of~$\sigma$, which is the
  second property of linearizability.

  We now show the first property of linearizability, i.e., that $\pi$ is
  also a view of $\sigma$ for all clients w.r.t. $G$.  The sequence~$\pi$
  is a view of $\sigma$ at a client~$C_i$ if it satisfies three conditions
  (Definition~\ref{def:view}).  The first conditions holds because $\pi$ is
  constructed as a permutation of $\sigma$.  Since each executed operation
  appears in $\sigma$ in terms of its invocation and response events, $\pi$
  contains all operations of all clients.  This implies the second
  condition of a view.  It remains to show that $\pi$ satisfies the
  sequential specification of~$G$.

  For reasoning about $G$, we introduce additional notation to capture the
  fact that it is not deterministic.  For a sequence $\omega$ of operations
  of~$G$ occurring in an actual execution, we write
  $\op{successful}(\omega)$ for the subsequence whose status was
  \str{success}, determined for each operation by the client that executed
  the operation.  Restricted to successful operations, $G$ is deterministic
  and reduces to~$F$.

  In particular, consider some operation~$o \in \pi$, executed by
  client~$C_i$ and fix a schedule that determines which operations are
  successful.  We want to show the following claim:
  \begin{quote}\em
    For any client $C_j$ (including $C_j = C_i$), the tuple $(s, r')
    \becomes F(s, o)$ computed when $C_j$ applies~$o$ in L\ref{c:doapply}
    satisfies:
    \begin{enumerate}\shortlist
    \item $(s, r') = F(s_0, \op{successful}(\pi|^o))$;
    \item If $o \in \op{successful}(\pi|^o)$, i.e., if $o$ is successful,
      then $r'$ is equal to the response~$r$ that $C_i$ has output when it
      completed~$o$ (L\ref{c:reply:end}); otherwise, $C_i$ has responded
      with~$\bot$
    \end{enumerate}
  \end{quote}
  We use induction on the operations sequence $\pi$ to show this.

  Consider the base case where $o$ is the first operation in~$\pi$ and
  recall that every client initializes its local state variable~$s$
  to~$s_0$.  Note that $S$ has not reported any pending operations to~$C_i$
  because $o$ is the first operation.  Thus, $C_i$ determines that the
  status of $o$ is \str{success}, computes $(s', r) \becomes F(s_0, o)$ and
  outputs~$r$.  When $C_j$ later receives $o$ in the \str{broadcast}
  message from~$S$ with sequence number~1, he applies $o$ because he learns
  status of $o$ in $z$.  Then $C_j$ updates the state~$s$ as
  $(s, r') \becomes F(s_0, o)$.  Since $F$ is deterministic,
  $(s,r') = (s',r)$ and the claim follows.

  Now consider the case when $o$ is not the first operation in $\pi$ and
  assume that the induction assumption holds for the operation that appears
  in $\pi$ before~$o$.
  If the status of $o$ is $\str{abort}$, then $o$ is filtered out by the
  $\op{successful}()$ operator in the claim; similarly, $C_j$ leaves the
  state $s$ unchanged upon applying~$o$
  (L\ref{c:ifapply}--L\ref{c:doapply}).  In addition, $C_i$ has responded
  with $\bot$ since $o$ was aborted
  (L\ref{c:conflict:no}--L\ref{c:conflict:end}).  The claim follows.

  Otherwise, if $o$ succeeds, we need to show that the state~$s$ at
  client~$C_j$ after applying~$o$ satisfies $(s, r)$ $=$ $F(s_0,
  \op{successful}(\pi|^o))$ and that the response of $C_i$ is~$r \neq
  \bot$.  Since~$S$ is correct, she assigns unique sequence numbers to the
  operations in the order in which she receives them in \str{invoke}
  messages (L\ref{s:invoke:begin}--L\ref{s:invoke:end}).  According to the
  code for confirming and applying operations, $C_i$ therefore processes
  (via L\ref{c:confirm}) a sequence of operations that is a prefix
  of~$\pi$, takes into account the status of each operation, and filters
  out those that abort~(L\ref{c:ifapply}--L\ref{c:doapply}).  This ensures
  the first property of the claim.

  Let $\rho$ be the sequence of operations that $C_i$ has confirmed before
  he received the \str{reply} containing~$o$; this sequence is in the order
  of the sequence numbers assigned by $S$ and in the order in which $C_i$
  confirmed these operations.  It follows from the construction of $\pi$
  that $\rho = \pi|^{o^*}$, where $o^*$ is the last operation in~$\rho$.
  The induction assumption implies that variable $s$ at $C_i$ after
  applying $o^*$ is equal to $s^*$, defined by 
  \begin{equation}\label{eq:sstar}
    (s^*, \cdot) = F(s_0,
    \op{successful}(\pi|^{o^*})) = F(s_0, \op{successful}(\rho)).
  \end{equation}
  Thus, $C_i$ starts processing the \str{reply} message for $o$ containing
  the list \var{Pend} from state~$s=s^*$
  (L\ref{c:reply:begin}--L\ref{c:reply:end}).
  $C_i$ constructs implicitly a permutation
  $\var{Pend-self} \concat \langle o \rangle \concat \var{Pend-other}$ of
  \var{Pend}.
  Recall that we consider the case where operation~$o$ succeeds and
  \var{Pend-other} does \emph{not} conflict with
  $\var{Pend-self} \concat \langle o \rangle$ in state~$s^*$, as ensured
  in~L\ref{c:conflict:no}.
  Thus, $C_i$ outputs response~$r$ given by
  $(\cdot, r) \becomes F(s^*, \var{Pend-self} \concat \langle o \rangle)$
  in L\ref{c:reply:end}.  The definition of non-conflicting operation
  sequences implies that $r$ is also equal to the response~$\bar{r}$ from
  \[
    (\cdot, \bar{r}) = F(s^*, \var{Pend-other} \concat \var{Pend-self}
    \concat \langle o \rangle)
  \]
  because this is a mixed operation sequence from $\var{Pend-other}$ and
  $\var{Pend-self} \concat \langle o \rangle$, which preserves the partial
  order of operations from the subsequences.

  It follows first from the construction of \var{Pend-other} and
  \var{Pend-self}, which contain all operations of \var{Pend} except for
  $o$ and the aborted pending-self operations of $C_i$, second, from their
  conflict-freedom, and, third, from recalling that \var{Pend-self} does
  not contain aborted pending-self operations that also the
  response~$\tilde{r}$ from
  \begin{equation}\label{eq:rtilde}
    (\cdot, \tilde{r}) = F(s^*, \op{successful}(\var{Pend}))
  \end{equation}
  satisfies $r = \bar{r} = \tilde{r}$.

  The definition of $F$ on operation sequences implies, furthermore, that
  \begin{equation}\label{eq:eqf}
    F(s^*, \op{successful}(\var{Pend})) = F(s_0, \op{successful}(\pi|^o)
  \end{equation}
  because $o$ is the last operation in~\var{Pend} and according to the
  definition of $s^*$ in~(\ref{eq:sstar}).

  To show that $r = r'$, where $r'$ is computed by $C_j$ when he
  applies~$o$ (L\ref{c:doapply}), note that $C_j$ has applied all
  successful operations in $\pi$ up to $o$ at this time and computed
  $(s, r') = F(s_0, \op{successful}(\pi|^o)$.  Combining this with
  (\ref{eq:eqf}) and (\ref{eq:rtilde}) now shows that $r' = r$ and the
  third property of a view follows.

  Note that the claim holds for any client~$C_j$, therefore, $\sigma$ is
  linearizable w.r.t.~$G$.
\end{proof}

\begin{lemma}
  If the clients execute all operations without overlaps, then all
  histories of \NAME are linearizable w.r.t.~$F$ and no operations abort.
\end{lemma}

\begin{proof}
  Consider an operation~$o$ that a client~$C_i$ has invoked and suppose
  towards a contradiction that it aborts.  According to the protocol, this
  occurs only if \var{Pend} in the \str{reply} message with~$o$ to $C_i$
  contains some pending-other operation, say, $o'$ executed by~$C_j$.  This
  implies that $o'$ has not been applied yet by $C_i$, even though $C_j$
  has applied it according to the assumption that the execution does not
  have any overlapping operations.

  However, because $C_j$ has applied $o'$ and $S$ is correct, it follows
  that $S$ has also sent the \str{broadcast} message containing $o'$ to
  $C_j$ earlier, before $C_j$ has applied~$o'$.  Note that messages between
  the correct server and one client are delivered in FIFO order.  Hence,
  $C_i$ receives the \str{broadcast} message corresponding to $o'$ and has
  applied $o'$ \emph{before} processing the \str{reply} message containing
  $o$.  This implies that $o' \not\in \var{Pending}$ according to server's
  operation (L\ref{s:invoke:begin}--L\ref{s:invoke:end}).  Hence, $C_i$
  does not abort, which contradicts the assumption.
\end{proof}

\subsection{The promised view of an operation}
\label{sec:promised}

In this and the next section, we prove Theorem~\ref{thm:forklin}.  The
proof starts by constructing a view for every client that includes all
operations that he has executed or applied, together with those of his
operations that some other clients have confirmed.  Since these operations
may have changed the state at other clients, they must be considered.  More
precisely, some $C_k$ may have confirmed an operation~$o$ executed by $C_i$
that $C_i$ has not yet confirmed or applied.  Then, in order to be
fork-linearizable even if $C_i$ will not confirm $o$ later, the view of
$C_i$ must include $o$ as well, including all operations that were
``promised'' to $C_i$ by $S$ in the sense that they were announced by $S$
as pending for~$o$.  It follows from the properties of the hash chain that
the view of $C_k$ up to $o$ is the same as $C_i$'s view including the
promised operations (Lemma~\ref{lem:promised}).  The view of $C_i$ further
includes all operations that $C_i$ has executed after~$o$.  Taken together
this will demonstrate that every execution of \NAME is fork-linearizable
w.r.t.~$G$ (Lemma~\ref{lem:view:history}).

Suppose a client~$C_i$ executes and thereby commits an operation~$o$.
We define the \emph{promised view to~$C_i$ of~$o$} as the sequence of
all operations that $C_i$ has confirmed before committing $o$,
concatenated with the sequence~$\var{Pend}$ of pending operations received
in the \str{reply} message during the execution of~$o$, including $o$
itself (according to the protocol $C_i$ verifies that the last
operation in~$\var{Pend}$ is $o$).

The protocol constructs a hash chain~$H$ over a sequence of (index,
operation, client)-triples of the form $(1, o_1, i_1), \dots, (l, o_l,
i_l)$.  Starting from $H[0] = \bot$, we set $H[k] \becomes
\op{hash}(H[k-1]\|o_k\|k\|i_k)$ for $k = 1, \dots, l$.  The value $h =
H[l]$ at the tip of the hash chain \emph{represents} the operation sequence
$\langle o_1, \dots, o_l \rangle$.  According to the collision-resistance
of the hash function, no two different operation sequences are represented
by the same hash value.

\begin{lemma}\label{lem:hashchain}
  After $C_j$ has confirmed some operation~$o$ at index~$l$, his hash-chain
  value $H[l]$ represents the sequence of operations that he has confirmed
  up to~$o$.
\end{lemma}

\begin{proof}
  Recall that $C_j$ extends $H$ in two places: when he confirms an
  operation at some index $l$ (L\ref{c:chain:2a}--L\ref{c:chain:2b}), and
  when he receives a \str{reply} message with pending operations
  (L\ref{c:chain:1a}--L\ref{c:chain:1b}).  According to the checks when
  $C_j$ receives an operation to confirm in a \str{broadcast} message
  (L\ref{c:confirm}), the client builds the hash chain $H$ incrementally,
  controlled by variable~$c$, in the sequence of the operations that he
  confirms.  An operation $o'$ from \var{Pend}, at some index $l'$ higher
  than $c$, might also have been inserted into $H$ within the loop
  (L\ref{c:loop:begin}--L\ref{c:loop:end}) earlier, when $C_j$ executes an
  operation of his own.  This is also controlled by~$c$ (L\ref{c:index}).
  But when $C_j$ later receives a \str{broadcast} message with this
  index~$l'$, any operation~$o^*$, and any hash-chain tip~$h$, he verifies
  the \str{commit}-signature (L\ref{c:confirm}) and checks that the
  hash-chain entry $H[l'] = \op{hash}(H[l'-1]\|o^*\|l'\|i)$ computed by
  himself is equal to the signed~$h$ (L\ref{c:chain:2test}).  Since this
  succeeds, $o' = o^*$ and $H[l]$ represents the sequence of operations
  that $C_j$ has confirmed up to~$o$.
\end{proof}

\begin{lemma}\label{lem:promised}
  If $C_j$ has confirmed some operation~$o$ that was committed by a
  client~$C_i$ (including $C_i = C_j$), then the sequence of operations
  that $C_j$ has confirmed up to (and including)~$o$ is equal to the
  promised view to $C_i$ of~$o$.  In particular:
  \begin{enumerate}\shortlist
  \item if clients $C_j$ and $C_k$ have confirmed an operation~$o$
    committed by $C_i$, then $C_j$ and $C_k$ have both confirmed the same
    sequence of operations up to~$o$;
  \item the promised view to $C_i$ of $o$ contains all operations
    executed by~$C_i$ up to~$o$.
  \end{enumerate}
\end{lemma}

\begin{proof}
  We first investigate the promised view to $C_i$ of $o$, which by
  definition consists of the sequence of operations that $C_i$ has
  confirmed, followed by the list $\var{Pend}$ in the \str{reply} message,
  including~$o$.  Consider the time when $C_i$ receives the \str{reply}
  message during the execution of~$o$.  We first show that when $C_i$
  commits $o$ with sequence number~$l$, the hash-chain entry $H[l]$
  represents the promised view to $C_i$ of $o$.  

  According to Lemma~\ref{lem:hashchain}, $H[c]$ represents the sequence of
  operations confirmed by $C_i$ so far.
  For every pending operation $p\in\var{Pend}$, client $C_i$ checks if he
  has already an entry in $H$ at index~$l$, which is the promised sequence
  number of~$p$ to $C_i$ according to~$\var{Pend}$.  If there is no such
  entry, he computes the hash value $H[l]$ as above.  Otherwise, $C_i$ must
  have received an operation for sequence number~$l$ earlier, and so he
  verifies that $o$ is the same pending operation as received before and
  stored in $H[l]$ (L\ref{c:chain:1test}).  Later, $C_i$ verifies that $o$
  itself has also been returned to him as pending~(L\ref{c:otest}).  Hence,
  the new hash value~$h$ stored in $H$ at the sequence number of $o$ (i.e.,
  $H[l]$ at L\ref{c:commit}) represents the promised view to~$C_i$ of~$o$.
  Then $C_i$ issues a \str{commit} signature~$\phi$ on $o$ and $h$ and
  sends $\phi$ to the server.

  When $C_j$ receives the \str{broadcast} message from~$S$ with the
  \str{commit}-signature $\phi$ of $C_i$ and operation~$o$ to be confirmed
  and applied by $C_j$ with sequence number~$l$, he verifies the
  \str{commit}-signature of $C_i$ on $o$, $l$, and $h$, and only confirms
  $o$ if the hash value satisfies~$h = H[l]$ (L\ref{c:chain:2test}).
  Recall from Lemma~\ref{lem:hashchain} that $H[l]$ represents the sequence
  of operations that $C_j$ has confirmed up to~$o$.  Noting that the hash
  function has no collisions, $h$ and $H[l]$ represent the same sequence of
  operations and the main statement of the lemma follows.

  The first additional claim follows by applying the lemma twice for $o$
  committed by~$C_i$.  For showing the second additional claim, we note
  that if $C_i$ confirms an operation by himself, then he has previously
  executed it.  There may be additional operations that $C_i$ has executed
  but not yet confirmed, but $C_i$ has verified according to the above
  argument that these were all contained in $\var{Pend}$ from the
  \str{reply} message.  Thus, they are also in the promised view of~$o$.
\end{proof}

\subsection{The view of a client}
\label{sec:view}
We construct a sequence~$\pi_i$ from $\sigma$ as follows.  Let $o$ be the
operation committed by $C_i$ which has the highest sequence number among
those operations of $C_i$ that have been confirmed by some client~$C_k$
(including $C_i$).  Let $\alpha_i$ be a sequence of operations constructed
as follows.  It contains all operations confirmed by $C_k$ up to and
including~$o$; if $C_i$ has confirmed $o$, then append the operations that
$C_i$ has confirmed after~$o$ (if any).

Furthermore, let $\beta_i$ be the sequence of operations committed by
$C_i$ with a sequence number higher than that of~$o$.  Then $\pi_i$ is
the concatenation of $\alpha_i$ and~$\beta_i$.  

Observe that by definition, every operation in $\alpha_i$ has been
confirmed by \emph{some} client and \emph{no} client has confirmed
operations from~$\beta_i$.

\begin{lemma}\label{lem:view}
  The sequence $\pi_i$ is a view of $\sigma$ at $C_i$ w.r.t.~$G$.
\end{lemma}

\begin{proof}
  Note that $\pi_i$ is defined through a sequence of operations that are
  contained in~$\sigma$.  Hence $\pi_i$ is sequential by construction.

  We now argue that all operations executed by $C_i$ are included
  in~$\pi_i$.  Recall that $\pi_i = \alpha_i \concat \beta_i$ and
  consider~$o$, the last operation of $C_i$ in~$\alpha_i$.  As $o$ has been
  confirmed by some $C_k$, Lemma~\ref{lem:promised} shows that $\alpha_i$
  is equal to the promised view to $C_i$ of $o$ and, furthermore, that it
  contains all operations that $C_i$ has executed up to~$o$.  By
  construction of~$\pi_i$ all other operations executed by~$C_i$ are
  contained in~$\beta_i$, and the property follows.

  The last property of a view requires that $\pi_i$ satisfies the
  sequential specification of~$G$.  Note that $G$ is not deterministic and
  some responses might be~$\bot$.  But when we ensure that two operation
  sequences of $G$ have responses equal to $\bot$ in exactly the same
  positions, then we can conclude that two equal operation sequences give
  the same resulting state and responses, from the fact that $F$ is
  deterministic.

  We first address the operations in~$\alpha_i$ and assume no operation
  aborts and returns~$\bot$.  Consider any $o_j \in \alpha_i$, executed by
  a client~$C_j$ (including $C_i = C_j$).  Lemma~\ref{lem:promised} implies
  that $\alpha_i|^{o_j}$ is a prefix of the promised view to $C_i$ of~$o$.
  We want to show that the response $r_j$ of $o_j$ to $C_j$ satisfies the
  specification of~$G$, i.e., that
  $(\cdot, r_j) = F(s_0, \op{successful} (\alpha_i|^{o_j}))$.

  For the point in time when $C_j$ executes $o_j$, define $\rho_j$ to be
  the sequence of operations that $C_j$ has confirmed prior to this and
  define $s_j$ to be the state resulting from applying the successful
  operations in~$\rho_j$, as stored in variable~$s$.  This implies that
  $(s_j, \cdot) = F(s_0, \op{successful}(\rho_j))$ according to the
  protocol (L\ref{c:bc:begin}--L\ref{c:bc:end}), and using the notation
  $\op{successful}(\cdot)$ from Lemma~\ref{lem:correct}.

  We want to show that the response of $o_j$ to $C_j$ satisfies $G$ as
  well.  Let $\var{Pend}$ be the pending operations contained in the
  \str{reply} message from~$S$ to $C_j$.  Observe that $C_j$ partitions
  $\var{Pend}$ into $\var{Pend-other}$ (pending-other operations),
  $\var{Pend-self}$ (successful pending-self operations), and the aborted
  pending-self operations of $C_j$, where $o_j$ is also among the
  pending-self operations.

  Client~$C_j$ then checks if \var{Pend-other} does not conflict with
  $\var{Pend-self} \concat \langle o_j \rangle$ in~$s_j$
  (L\ref{c:conflict:no}), and if this is the case (L\ref{c:simulate}),
  ~$C_j$ computes the response~$r_j$ for $o_j$ from state~$s_j$ as $(\cdot,
  r_j) \becomes F(s_j, \var{Pend-self} \concat \langle o_j \rangle)$.
  Since $\var{Pend-self}$ and $\var{Pend-other}$ preserve the relative
  order of operations in $\var{Pend}$, the definition of non-conflicting
  operations implies that the responses of $C_j$ from $F(s_j,
  \op{successful} (\var{Pend}))$ and $F(s_j, \var{Pend-self} \concat
  \langle o \rangle)$ are equal.
  This demonstrates that $(\cdot, r_j)$
  $=$ $F(s_j, \op{successful} (\var{Pend}))$
  $=$ $F(s_0, \op{successful} (\alpha_i|^{o_j}))$, i.e., that 
  $o_j$ satisfies the sequential specification of $G$ assuming no aborts.
  Recall this holds for any operation $o_j$ in $\alpha_i$ and that
  $\pi_i = \alpha_i \concat \beta_i$.

  To conclude the argument, we still have to show that the abort status for
  every operation $o_j \in \alpha_i$ is the same for any client $C_k$
  (including $C_k = C_i$) who confirms~$o_j$, and $C_j$ (who has
  committed~$o_j$).  Then they will produce the same responses and same
  state.  Note that when $C_j$ executes $o_j$, he either computes a
  response according to~$F$ or aborts the operation, declaring its status
  to be \str{success} or \str{abort}, respectively
  (L\ref{c:conflict:no}--L\ref{c:conflict:end}).  The status~$z$ is signed,
  sent to~$S$ in the \str{commit} message, and should be received in the
  \str{broadcast} message (L\ref{c:bc:begin}) by~$C_k$.  Since $C_k$ has
  confirmed $o_j$, he has verified the \str{commit} signature and this
  implies that the status taken into account by $C_k$ is also equal to~$z$,
  as used to determine whether he updates the state with~$o_j$
  (L\ref{c:confirm}--L\ref{c:doapply}).

  Furthermore, we need to show that the operations in $\beta_i$ satisfy the
  specification of~$G$, where $\beta_i$ consists of operations committed by
  $C_i$ with a sequence number higher than that of~$o$.  According to the
  earlier argument about $\alpha_i$ and considering that $C_i$ has
  confirmed $o$ when computing the responses of operations in $\beta_i$,
  when $C_i$ receives the \str{reply} message for~$o$
  (L\ref{c:reply:begin}), its state $s$ results from confirming and
  applying all operations in $\alpha_i$.  Hence,
  $(s, \cdot) = F(s_0, \op{successful}(\alpha_i))$.  For every successful
  $o_i \in \beta_i$ client~$C_i$ computes the response~$r_i$ of $o_i$
  (L\ref{c:simulate}) as
  $(\cdot, r_i) = F(s, \var{Pend-self} \concat \langle o_i \rangle)$, where
  it is easy to see that the variable \var{Pend-self} (which does not
  contain aborted operations) is a prefix of $\op{successful}(\beta_i)$.
  Since $C_i$ executes and commits the operations of $\beta_i$ in the order
  of their sequence numbers, it follows that also
  $(\cdot, r_i) = F(s, \op{successful}(\beta_i|^{o_i}))$ and this implies
  $(\cdot, r_i) = F(s_0, \op{successful}(\alpha_i \concat \beta_i|^{o_i}))$
  by the definition of~$F$.  Thus, all operations in $\beta_i$ satisfy the
  specification~$G$ as well.
\end{proof}

\begin{lemma}\label{lem:app:realtime}
  If some client $C_j$ confirms an operation~$o_1$ before an
  operation~$o_2$, then $o_2$ does not precede $o_1$ in the execution
  history~$\sigma$.
\end{lemma}

\begin{proof}
  Let $\mu_j$ denote the sequence of operations that $C_j$ has confirmed up
  to~$o_2$.  According to the protocol logic
  (L\ref{c:bc:begin}--L\ref{c:bc:end}), $\mu_j$ contains $o_1$, and $o_1$
  has a smaller sequence number than~$o_2$.  Suppose $o_2$ was executed
  by~$C_i$.  Lemma~\ref{lem:promised} shows that $\mu_j$ is equal to the
  promised view to $C_i$ of $o_2$, hence, $o_1$ is contained in the
  promised view to $C_i$ of~$o_2$.  If $C_i$ has confirmed $o_1$ earlier,
  then $o_2$ does not precede $o_1$.  If $o_1$ is pending for $o_2$, then
  $o_1$ has been invoked by a client before $o_2$, as validated by~$C_i$
  through verifying the corresponding \str{invoke}
  signature~(L\ref{c:signinvoke}).  Since this occurs before $o_2$
  completes, $o_1$ has been invoked before $o_2$ completed.
\end{proof}

\begin{lemma}\label{lem:view:realtime}
  The sequence~$\pi_i$ preserves the real-time order of~$\sigma$.
\end{lemma}

\begin{proof}
  Recall that $\pi_i = \alpha_i \concat \beta_i$ and consider first the
  operations in~$\alpha_i$, which have been confirmed by some client.
  Lemma~\ref{lem:app:realtime} shows that these operations preserve the
  real-time order of~$\sigma$.  Second, the operations in $\beta_i$ are
  ordered according to their sequence number and they were committed
  by~$C_i$.  According to the protocol, $C_i$ executes only one operation
  at a time and always assigns a sequence number that is higher than the
  previous one.  Hence, $\beta_i$ also preserves the real-time order
  of~$\sigma$.

  We are left to show that no operation in $\beta_i$ precedes an operation
  from $\alpha_i$ in~$\sigma$.  Recall that $\alpha_i$ consists of
  operations that have been confirmed and $\beta_i$ are operations executed
  and committed by~$C_i$.  Let~$\tilde{o}$ be the last operation
  of~$\alpha_i$ and suppose it has sequence number~$l$.
  
  If some $C_k \neq C_i$ has confirmed $\tilde{o}$, then $\tilde{o}$ has
  been executed by~$C_i$ according to the definition of $\alpha_i$, and
  $\tilde{o}$ has already completed before $C_i$ invokes the first
  operation of $\beta_i$ (which have sequence numbers larger than~$l$),
  according to the assumption that $\sigma$ is well-formed.

  Otherwise, $C_i$ has confirmed $\tilde{o}$ and some $C_j \neq C_i$ has
  executed $\tilde{o}$.  Consider the time when $C_i$ confirms~$\tilde{o}$:
  $C_j$ must have already completed~$\tilde{o}$ because $C_i$ verified the
  \str{commit} signature for~$\tilde{o}$ issued by $C_j$ when $\tilde{o}$
  completed.  Since $C_i$ has verified this signature for
  confirming~$\tilde{o}$ (L\ref{c:confirm}), $C_i$'s local sequence-number
  variable~$c$ is at least~$l$ at that time.  As operations of~$\beta_i$
  have larger sequence numbers than~$l$ by definition, the protocol for
  handling \str{reply} messages (L\ref{c:reply:begin}--L\ref{c:reply:end})
  implies that all those operations were invoked after $\tilde{o}$
  completed.
\end{proof}

\begin{lemma}\label{lem:easy}
  If some operation $o_j \in \pi_i$ executed by $C_j$ has been confirmed by
  a client~$C_k$ (including $C_i$), then $o_j \in \alpha_i$ and
  $\alpha_i|^{o_j} = \pi_i|^{o_j}$; furthermore, $\alpha_i|^{o_j}$ is equal
  to the promised view to $C_j$ of~$o_j$.
\end{lemma}

\begin{proof}
  Consider the case that $C_k = C_i$ has confirmed $o_j$.  Then $o_j \in
  \alpha_i$ according to the definition of $\alpha_i$.  The second
  statement is an immediate consequence of Lemma~\ref{lem:promised}, since
  $C_i$ has confirmed~$o_j$.

  Otherwise, some $C_k \neq C_i$ has confirmed $o_j$.  If $C_j = C_i$, then
  $o_j \in \alpha_i$ by definition since $o_j$ has been confirmed. 
  If $C_j \neq C_i$, then $o_j \in \alpha_i$ because $o_j \in \pi_i$ but
  $\beta_i = \pi_i \setminus \alpha_i$ contains only operations executed
  by~$C_i$.  The second statement is an immediate consequence of
  Lemma~\ref{lem:promised}, since $C_k$ has confirmed~$o_j$ and $\alpha_i$
  is defined accordingly.
\end{proof}

\begin{lemma}\label{lem:view:history}
  If $o \in \pi_i \cap \pi_j$ then $\pi_i|^o = \pi_j|^o$.
\end{lemma}

\begin{proof}
  As $\pi_i = \alpha_i \concat \beta_i$ and $\pi_j = \alpha_j \concat
  \beta_j$, we need to consider four cases to analyze all operations
  that can appear in $\pi_i \cap \pi_j$ and the rest are symmetrical.

  \begin{enumerate}\shortlist
  \item $o \in \alpha_i$ and $o \in \alpha_j$: This implies that $o$ has
    been confirmed.  Lemma~\ref{lem:easy} implies that
    $\alpha_i|^o = \alpha_j|^o$.

  \item $o \in \beta_i$ and $o \in \alpha_j$: This case cannot occur,
    since no client has confirmed operations from $\beta_i$ by
    definition.

  \item $o \in \alpha_i$ and $o \in \beta_j$: Analogous to the case
    above.

  \item $o \in \beta_i$ and $o \in \beta_j$: This case cannot occur,
    since $\beta_i$ and $\beta_j$ contain only pending-self operations
    of $C_i$ and $C_j$, correspondingly.
\end{enumerate}
\end{proof}

\section{Conclusion}
\label{sec:conclusion}

This paper has presented \NAME, the Conflict-free Operation verification
Protocol, which lets a group of clients execute a generic service
coordinated by a remote, but untrusted server.  \NAME ensures \forklin and
allows clients to easily verify the consistency and integrity of the
service responses.  In contrast to previous work, \NAME is wait-free and
supports non-con\-flic\-ting operation sequences (but may sometimes abort
conflicting operations);

In \NAME every client executes all operations of the common service and
maintains the state, similar to a replicated state machine~\cite{schnei90}.
It is possible to improve the efficiency of \NAME for specific services
that permit efficient authentication of remote state, in order to reduce
the work of the clients and to keep the (potentially large) state only at
the server.  This goal can typically be achieved for functionalities that
support authenticated data structures~\cite{tamass03}.  In a successor to
this work, Brandenburger et al.~\cite{brcakn17} show how apply this method
to protect the integrity and consistency of data in a cloud object store.

Efficient authenticated data structures are only available for certain
functionalities.  Therefore, an important direction for future work lies in
combining generic protocols for cryptographically verifiable
computation~\cite{walblu15} with \NAME, to reduce the client workload for
arbitrary computations and to guarantee integrity and consistency with
\forklin semantics to multiple clients.

\section*{Acknowledgments}

We thank Marcus Brandenburger for interesting discussions and valuable
comments.  We are grateful to the anonymous reviewers for important and
constructive comments.

This work has been supported in part by the European Union's Seventh
Framework Programme (FP7/2007--2013) under grant agreement number
ICT-257243 TCLOUDS; in part by the European Commission through the Horizon
2020 Framework Programme (H2020-ICT-2014-1) under grant agreements number
644371~WITDOM and 644579~ESCUDO-CLOUD; and in part by the Swiss State
Secretariat for Education, Research and Innovation (SERI) under contracts
number 15.0098 and 15.0087.

\bibliographystyle{abbrv}
\bibliography{flc}

\end{document}